# 2D Weak Localization in Trilayer Ruddelsden-Popper Nickelates

Hyo-Bin Ahn[&], Xinglong Chen[#], Yu Zhang, Hong Zheng, Michael R. Norman, J. F. Mitchell, Daniel Phelan, and Ulrich Welp[&]

Material Science Division, Argonne National Laboratory, Lemont, Illinois 60439, USA

Ruddlesden-Popper (RP) nickelates superconduct when pressure suppresses intertwined charge- and spin-density waves. Here, we employ measurements of quantum corrections to magnetoconductivity as a probe of the dimensionality, phase coherence, and scattering mechanisms of the underlying single crystal trilayer RP nickelate $Pr_4Ni_3O_{10}$ and $La_4Ni_3O_{10}$. We observe signatures of 2D Weak Localization (WL), implying that the in-plane electron transport is in the quantum diffusive regime, while the out-of-plane transport is incoherent, and the electronic ground state is a quasi-2D disordered Fermi liquid. Application of pressure up to 3 GPa continuously suppresses signatures of WL, potentially signaling a 2D/3D dimensional crossover and an eventual pressure-driven transition into a superconductor.

The recent observations of superconductivity under high pressure in Ruddlesden-Popper nickelates with critical temperatures approaching 100 K [1-4] are noteworthy because superconductivity in these transition metal oxides apparently defies the standard cuprate paradigm of a hole-doped Mott insulator with $3d^9$ configuration [5-8]. Instead, superconductivity occurs in bilayer $R_3Ni_2O_7$ and trilayer $R_4Ni_3O_{10}$ compounds where the electron counts are nominally $3d^{7.5}$ and $3d^{7.33}$, respectively. Under ambient pressure, strongly coupled charge density wave (CDW) and spin density wave (SDW) (in the following denoted collectively as DW) states have been identified in the trilayer materials [9, 10]. Pressure suppresses the DW and induces superconductivity, suggestive of competition between these two quantum states, which is reminiscent of Fe-based, Kagome and cuprate superconductors. Understanding the nature of the charge transport and electron scattering can provide insight into the complex interplay among electron itineracy, spin interactions, superconductivity, and localization in these materials.

We have carried out the first (to our knowledge) study of quantum corrections to the magnetoconductivity on single crystals of the trilayer nickelates $Pr_4Ni_3O_{10}$ (PNO) and $La_4Ni_3O_{10}$ (LNO) at ambient pressure and under pressures up to 3 GPa. At low temperatures, a negative magnetoresistance (NMR) arises at low magnetic fields consistent with 2D Weak Localization (WL). Surprisingly, WL (as opposed to antilocalization) is found despite the high atomic numbers (*i.e.*, strong spin-orbit scattering) of the constituent La/Pr atoms. Likewise, quantum phase coherence occurs despite the presence of magnetic Ni and Pr atoms. We show that the robust WL results from (1) the partial cancellation of effects arising from spin-orbit and spin-flip scattering and (2) the spatial isolation of the rare earth ions from the $NiO_2$-layers, which minimizes the effects of spin-orbit and spin-flip scattering. The NMR displays 2D scaling with field orientation, demonstrating its orbital nature. Furthermore, WL is progressively suppressed under pressure up to 3 GPa, indicating a possible 2D/3D crossover and providing a link between the ambient-pressure normal state and the high-pressure superconducting state, whose intermediate transport regime remains poorly understood. Our results demonstrate that in-plane electron transport is in the quantum diffusive regime and exhibits remarkably similar characteristics for materials with different crystallographic, electronic and magnetic backgrounds. At the same time, out-of-plane transport is incoherent, implying a quasi-2D disordered Fermi liquid ground state.

Single crystals of PNO and LNO were synthesized using a high-pressure floating-zone furnace [11].. Comparing samples containing magnetic/non-magnetic ions on the rare earth site aids in disentangling the effects of different scattering mechanisms, *i.e.*, spin-orbit and spin-flip. The magnetic characteristics of the PNO and LNO samples are presented in the Supplemental Materials [12].

Fig. 1 shows the temperature dependence of the in-plane resistivity of three PNO crystals, labeled S1, S2 and S3, in panel (a), while panel (b) displays data for two LNO crystals. Clear step-like anomalies near $T_{\mathrm{DW}}$ of 158 K and 135 K indicate the DW transition in PNO and LNO, respectively, consistent with earlier reports [11, 13, 14]. With decreasing the temperature, the resistivity of all samples displays metallic characteristics (*i.e.*, $\mathrm{d}\rho/\mathrm{d}T > 0$) with temperature dependences that are approximately linear above the DW transition, with slopes ranging from 1 to 2 μΩ cm/K. These values agree well with those compiled for a series of materials displaying strange metal behavior [15]. The resistivity of some samples displays an upturn upon cooling to the lowest temperatures, e.g., PNO-S1, while for others, it nearly saturates (PNO-S2, S3). As shown in the inset of Fig. 1, the low temperature resistivity for the samples that show an upturn can reasonably be described by $\rho = \rho_0 - \beta T^{1/2}$. A similar temperature dependence has been reported for polycrystalline $R_4Ni_3O_{10}$ [14, 16-18].

The theoretical analysis [19] of electric transport in disordered materials reveals that WL and weak anti-localization (WAL) induce corrections to the resistivity that are proportional to $T^{1/2}$ for a 3D system and proportional to $ln(T)$ in 2D. However, the experimental determination of these quantum corrections requires knowledge of the 'normal' baseline $T$-dependence, which currently is poorly understood and sample dependent (see Fig. 1 and Fig. S1 in Supplemental Materials [12]). Therefore, the low-$T$ variation of resistivity alone is not a reliable indicator for the dimensionality of quantum transport [20]. In the following, we show that the magnetoresistance (MR) and its angular dependence unambiguously establish that the quantum corrections in single-crystal LNO and PNO are 2D.

The MR of PNO-S1, defined as MR = $(\rho(B)-\rho(0))/\rho(0)$, and LNO-S1 is shown for various temperatures in Figs. 2(a, b). An unexpected feature revealed by these data is the emergence of a cusp-like, NMR in fields below ~ 2 T and at temperatures below 30 K. This feature is superimposed upon a quadratic, positive MR (PMR) that dominates at high fields. At higher temperatures, the NMR disappears, and only the positive quadratic PMR remains. The PMR is small (i.e., less than ~0.5% in 9 T) and continuously diminishes with further increasing temperature. Such behavior is expected in a Drude-type description of metallic transport.

The observation of similar NMR in LNO without local 4*f*-electron moments and in PNO with local 4*f*-moments rules out spin-flip scattering by the 4*f*-moments as the underlying mechanism. The data shown in Fig. 2 do, however, display the typical signatures of quantum corrections to the conductivity arising in disordered systems [19, 21-24]. In systems for which the inelastic electron mean-free path, $l_\varphi$, is much longer than the elastic mean-free path, $l_e$, electrons can maintain phase coherence even after undergoing multiple elastic scattering events. As a result, quantum corrections to the conductivity can arise in the form of WL and WAL. In the former, constructive interference between time-reversed electron paths leads to an increase in resistance at low temperatures and NMR, while in the latter destructive interference causes a reduction of the low-$T$ resistance and PMR. Extensive theoretical work has established the temperature and magnetic field dependence of conductivity due to localization effects in 2D [21, 25-29] and in 3D systems [30-34].

We found that the data in Fig. 2 are incompatible with the 3D formalism, strongly suggesting a 2D description. The quantum corrections to the conductivity in 2D can be cast in the form [21, 27, 35]:

$$\Delta\sigma(H) = \sigma(H) - \sigma(0) = \frac{\alpha e^2}{2\pi^2\hbar}\left[\frac{3}{2}F\left(\frac{H_\varphi^* + 4/3H_{so}^*}{H}\right) - \frac{1}{2}F\left(\frac{H_\varphi^*}{H}\right)\right] \quad (1)$$

where $F(x) = \Psi(1/2 + x) - ln(x)$ with $\Psi$ the digamma function. $H_\varphi^* = H_\varphi + 2H_s$ and $H_{so}^* = H_{so} - H_s$ are the effective dephasing and spin orbit fields given by the combined effects of inelastic, spin-flip and spin-orbit scattering, respectively, with $H_{so}$, $H_s$ and $H_\varphi$ the characteristic fields describing spin-orbit, spin-flip and inelastic scattering. These fields are related to the corresponding scattering rates $1/\tau_k$ and length scales $l_k$ through $H_k = \hbar/4eD\tau_k = \hbar/4el_k^2$. $D$ is the electron diffusion coefficient of the elastically scattered electrons, and $\alpha$ describes the overall scale of the correction. Our data clearly reveal WL (as opposed to WAL), implying that spin-orbit scattering is comparatively weak [21]. In fact, the condition for seeing a NMR, i.e., WL, at low fields is $H_{so}^*/H_\varphi^* < 9/16$. At the same time, the data shows a relatively low scale for the dephasing field of $\mu_0 H_\varphi^* \sim 0.03$ T at low temperatures (see below), placing an upper bound on $H_s$. Furthermore, as is indicated by the expression for $H_{so}^*$, the effects of spin-orbit and magnetic scattering partially cancel each other. We therefore assume that $H_{so}^*$ can be neglected in comparison to $H_\varphi^*$ thereby recovering the commonly used simplified expression in terms of the effective dephasing field $H_\varphi$ [27]:

$$\Delta\sigma(H) = \frac{\alpha e^2}{4\pi^2\hbar}\left[\Psi\left(\frac{1}{2}+\frac{H_\varphi}{H}\right) - ln\left(\frac{H_\varphi}{H}\right)\right] + \beta H^2 \quad (2)$$

We include a quadratic term $\beta H^2$ to account for the Drude-like background magnetoconductivity. More details of the fitting procedure are presented in the Supplementary Materials section [12].

As shown in Figs. 2(a, b), fits according to Eq. (2) describes the data well. Fig. 2(c) displays the temperature dependence of the resulting phase coherence length $1/l_\varphi^2 = 4eH_\varphi/\hbar$, which is proportional to the scattering rate $1/\tau_\varphi$. $l_\varphi^{-2}$ decreases upon cooling; the expected behavior as the probability of inelastic scattering decreases at low temperatures [23]. At temperatures above ~8 K, the temperature variation is found to be $\sim T^{1.1}$ (PNO) and $\sim T^{0.9}$ (LNO), respectively, which is close to the linear behavior that is characteristic for predominant inelastic electron-electron scattering in 2D [23, 36]. Below ~8 K, the apparent levelling-off in scattering rate can be attributed to residual inelastic scattering [23].

To confirm 2D quantum transport in PNO, we show angle-dependent magneto-transport measurements in Fig. 3(a). In these measurements, the magnetic field and current directions are kept perpendicular to each other, as indicated in the inset. As the field turns into the in-plane direction, the positive curvature of the MR is gradually suppressed, while the field range where NMR appears broadens. When the field is aligned close to the in-plane, the NMR dominates in the entire field range up to 14 T. A study of the in-plane magnetoconductivity that includes spin-orbit, spin-flip scattering and Zeeman effect [28] finds the MR to be negative for $1/\tau_s > 1/\tau_{so}$. Thus, these data imply that $H_{so}^*$ is in fact negative and thereby promotes the observation of WL. Fig. 3(b) shows the MR data from Fig. 3(a) plotted versus the out-of-plane component of the magnetic field ($B$ sin($\theta$)) for various angles. Remarkably, all MR traces scale to a single curve at low fields in the range of NMR, corroborating that the NMR of PNO is a 2D phenomenon, depending only on the out-of-plane field component. This result also implies that the MR does not derive from a Kondo effect, which is expected to be isotropic.

Typically, 2D behavior of quantum corrections arises in thin-film samples when the phase-coherence length, $l_\varphi$, exceeds the sample thickness. Examples include the observation of 2D WL in 7-unit cell thick $LaNiO_3$ films [37], while measurements on 240-nm thick epitaxial films [20] reveal 3D behavior. Thus, single crystal samples are highly desirable to establish the *intrinsic* dimensionality of the quantum corrections in highly anisotropic materials such as layered materials, in which a dimensional crossover from 3D to 2D quantum corrections can occur [38, 39]. The criterion [39] defining the crossover is given as $t \lesssim h/\tau$, implying that the elastic mean free path in the perpendicular direction is of the order of the interlayer spacing. Here, $1/\tau$ is the total scattering rate and $t$ is the hopping energy in the perpendicular direction. At this cross over, $c$-axis transport becomes incoherent and is no longer described by quasi-classical Boltzmann theory [40]. Instead, in the optical conductivity, spectral weight is shifted from the (metallic) Drude peak to high energies thereby effectively reducing the dc-conductivity. This process is promoted by strong electron-electron correlations that drive the system towards a Mott insulating state.

Although reported values of the resistivity of single-crystal PNO and LNO vary substantially [11, 13, 41, 42], it is evident that these materials are highly resistive. Specifically, the $c$-axis resistivity far exceeds typical estimates of the Mott-Ioffe-Regel limit, consistent with the notion of incoherent $c$-axis transport, see the data in the inset of Fig. 1(a) [40, 43]. Furthermore, recent optical reflectivity measurements on LNO [44, 45] indicate the loss of the Drude peak for $c$-axis polarized light at low temperatures, implying an enormous apparent anisotropy of the dc-conductivity of ~366 [44] and ~2600 [45], that emerges below the DW transition. These findings are also in qualitative agreement with ARPES measurements [42], further supporting the scenario of incoherent charge dynamics along the *c*-axis. While the crystal structure is inherently quasi-2D due to the NiO-trilayers being separated via rock salt layers along with an in-plane lattice translation, the *c*-axis decoupling is further enhanced by a spin-density wave-induced redistribution of the $Ni\ 3d_{3z^2-r^2}$ orbital occupation at the DW transition [45].

Magneto-transport measurements under hydrostatic pressures of up to 3 GPa. (Fig. 4) reveal an overall resistivity decrease with increasing pressure, while both the DW transition and the WL contribution are progressively suppressed, as shown in Fig. 4(b) [41, 46-48]. The pressure induced suppression of WL could arise for several reasons: With pressure, the resistivity decreases implying an increase of the mean free path. This results in a reduced probability of forming interfering time-reversed electron trajectories, thus reduced WL/WAL effects. Second, with increasing pressure, bandwidths and interlayer hopping amplitudes increase [49-52] thereby weakening the quasi-2D nature of the electronic transport and promoting a crossover towards a more 3D electronic state. Generally, WL/WAL effects are weaker in three dimensions [19], therefore, the pressure evolution shown in Fig. 4 is consistent with the approach to a dimensional crossover. The low-pressure data are well described in the 2D framework as described in the manuscript. In contrast, the data at 2–3 GPa can be fit with both the 2D and the 3D form, possibly signaling a dimensional crossover. However, owing to the small size of the effect, the fits carry a high degree of uncertainty. We note that pressure-induced interlayer coupling ultimately leads to a remarkably isotropic superconducting state in PNO and LNO [46, 55].

It is astounding that we observe clear WL with relatively long phase-coherence lengths in a material containing high concentrations of both magnetic and heavy ions. Although sizable induced magnetic moments approaching the full free-ion effective value reside on the Pr ions (see Fig. S2 in Supplemental Materials [12]), they are located between the $NiO_2$ layers, resulting in spatial isolation from the conduction electrons, which are largely confined to the $NiO_2$-planes. A similar scenario arises in the cuprate superconductor $YBa_2Cu_3O_7$ where most rare earth substitutions on the Y-site have little effect on the superconducting state [53]. Similarly, the overlap between the itinerant electron wavefunctions and the Pr 4*f* magnetic moments in PNO is weak, leading to a reduced spin-flip scattering rate, and therefore limited dephasing of quantum interference. This reasoning is supported by our experimental observation of similar quantum corrections in the Pr and the La compound showing that the Pr-moments are decoupled from *ab*-plane transport. Furthermore, the *c*-axis resistivity data on PNO shown in Fig. 1(a) reveals an anomaly around 25 K that is not present in LNO data [45] and marks the onset of magnetic order on the $Pr_A$ (rock salt) sites [10]. This anomaly is entirely absent in *ab*-plane transport, directly showing that in-plane transport does not probe or couple to the rare earth ions. We surmise that any residual spin-flip scattering arises primarily due to the Ni moments. However, at temperatures below $T_{DW}$, the Ni moments are ordered in a SDW state thereby reducing the spin-flip scattering rate at low temperatures.

In this study, we investigated the electrical and magnetic properties of single-crystal $Pr_4Ni_3O_{10}$ and $La_4Ni_3O_{10}$. We have observed 2D WL at low temperatures in PNO and LNO; this observation is important because it establishes the 2D quantum transport characteristics, the diffusive nature of the in-plane transport as well as incoherent out-of-plane transport. The observation of WL in a material containing high concentrations of heavy elements and magnetic moments is unexpected. We attribute our findings to the partial cancellation between the effects of spin-orbit and spin-flip scattering, as well as to the spatial isolation of the rare earth ions from the $NiO_2$-planes; these ions are largely decoupled from electron transport. The angular dependence of the MR establishes the 2D nature and orbital origin of the effect; the Kondo contribution is negligible in the observed phenomena. The quasi-2D nature of the transport is in support of those models [54] where the $Ni\ 3d_{x^2-y^2}$ orbitals dominate the low energy physics. This occurs because the density wave gap primarily affects the $Ni\ 3d_{3z^2-r^2}$ orbitals, giving rise to incoherent transport along the c-axis, but in-plane metallic transport dominated by the $Ni\ 3d_{x^2-y^2}$ orbitals [42, 44, 45]. Under pressure, the density wave is suppressed and the $Ni\ 3d_{3z^2-r^2}$ orbitals come back into play. This gives rises to a coherent Fermi liquid with a superconducting phase which is remarkably isotropic [46, 55]. In this sense, our findings shed light on the transport characteristics of the parent state of the high-pressure superconductor, highlighting possible prerequisites for the emergence of superconductivity in trilayer nickelates.

**Acknowledgements**

This work was supported by the U. S. Department of Energy, Office of Science, Basic Energy Sciences, Materials Sciences and Engineering Division. Work performed at the Center for Nanoscale Materials, a U.S. Department of Energy Office of Science User Facility, was supported by the U.S. DOE, Office of Basic Energy Sciences, under Contract No. DE-AC02-06CH11357.

**References**

& corresponding authors: ahyobin@anl.gov; welp@anl.gov

# current address: School of Physics, Southeast University, Nanjing 211189, China.

## Figures

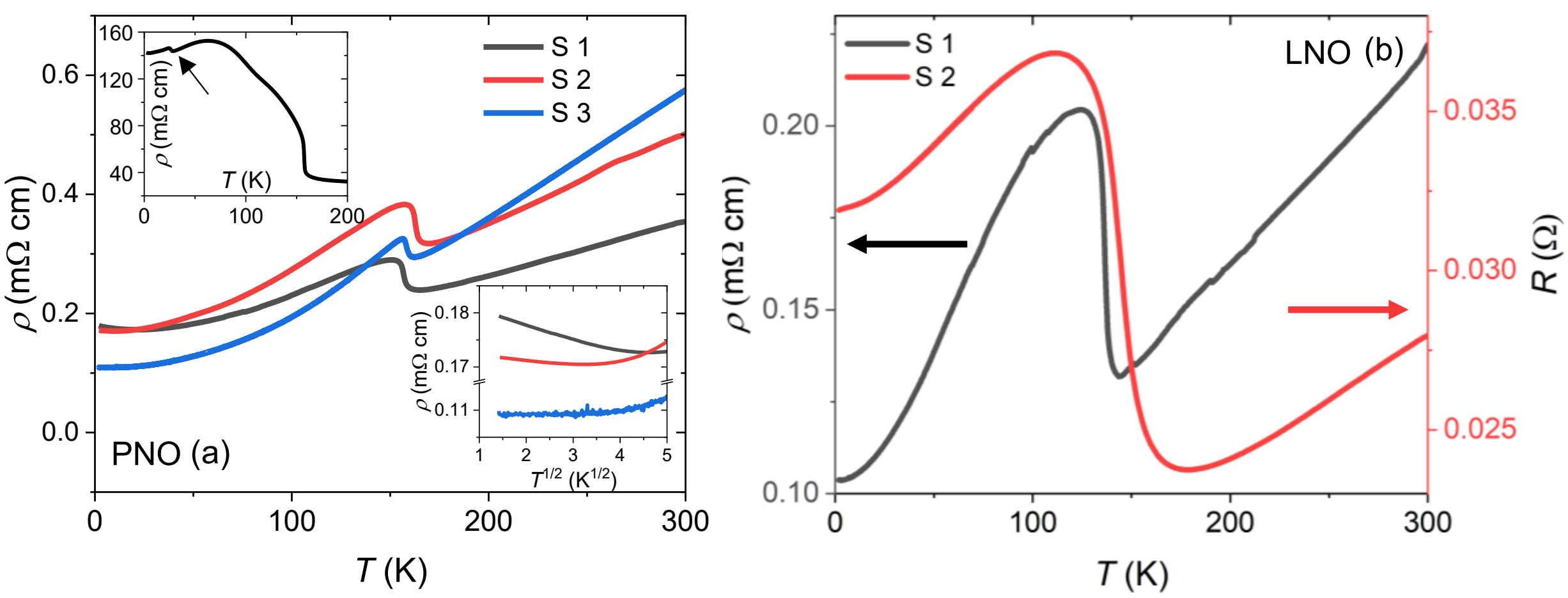

Fig. 1: Temperature dependence of the in-plane resistivity of $Pr_4Ni_3O_{10}$ crystals S1, S2, and S3 (a) and $La_4Ni_3O_{10}$ crystals S1 and S2 (b). The upper left inset shows the temperature dependence of the *c*-axis resistivity of $Pr_4Ni_3O_{10}$. The arrow marks the anomaly due to the onset of spin-density wave order on the $Pr_A$ (rocksalt) sites. The bottom right inset shows the low-temperature upturn of the resistivity plotted as a function of $T^{1/2}$.

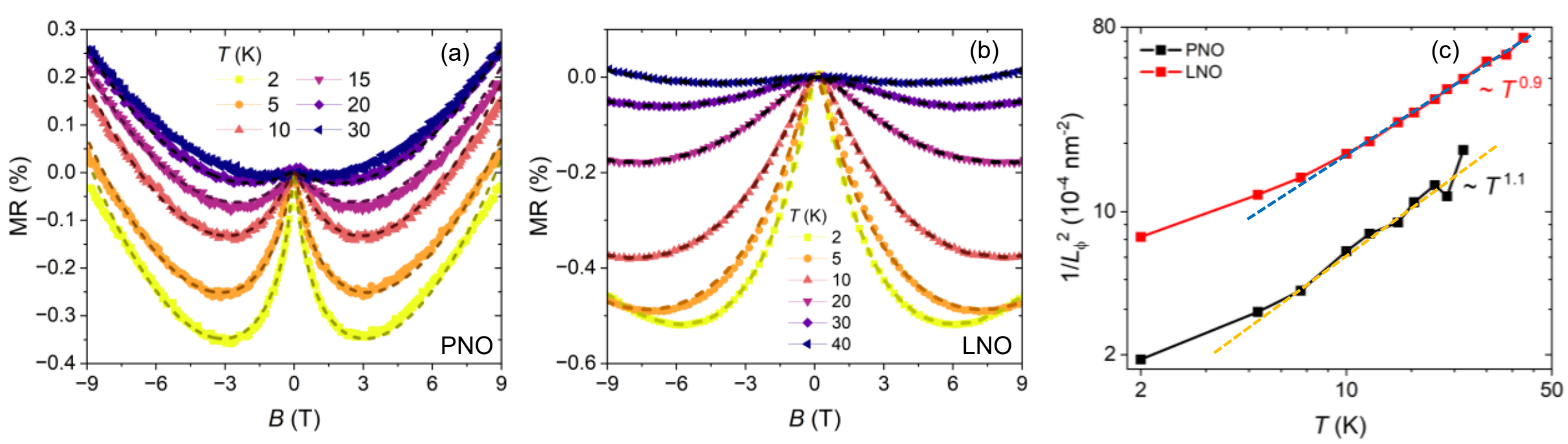

Fig. 2: MR of sample $Pr_4Ni_3O_{10}$-S1 (a) and of sample $La_4Ni_3O_{10}$-S2 (b) measured at various temperatures with $B//c$ while current flows through in-plane direction. The dotted lines represent fits according to Eq. (2). (c) Temperature dependence of the square of the inverse phase coherence length $1/l_\varphi^2$ of $Pr_4Ni_3O_{10}$ and $La_4Ni_3O_{10}$ plotted on a log-log scale. The dashed lines indicate a power-law variation of $T^{1.1}$ and $T^{0.9}$, respectively. For reference, the phase coherence field $H_\varphi$ of $Pr_4Ni_3O_{10}$ at 2 K is ~0.03 T.

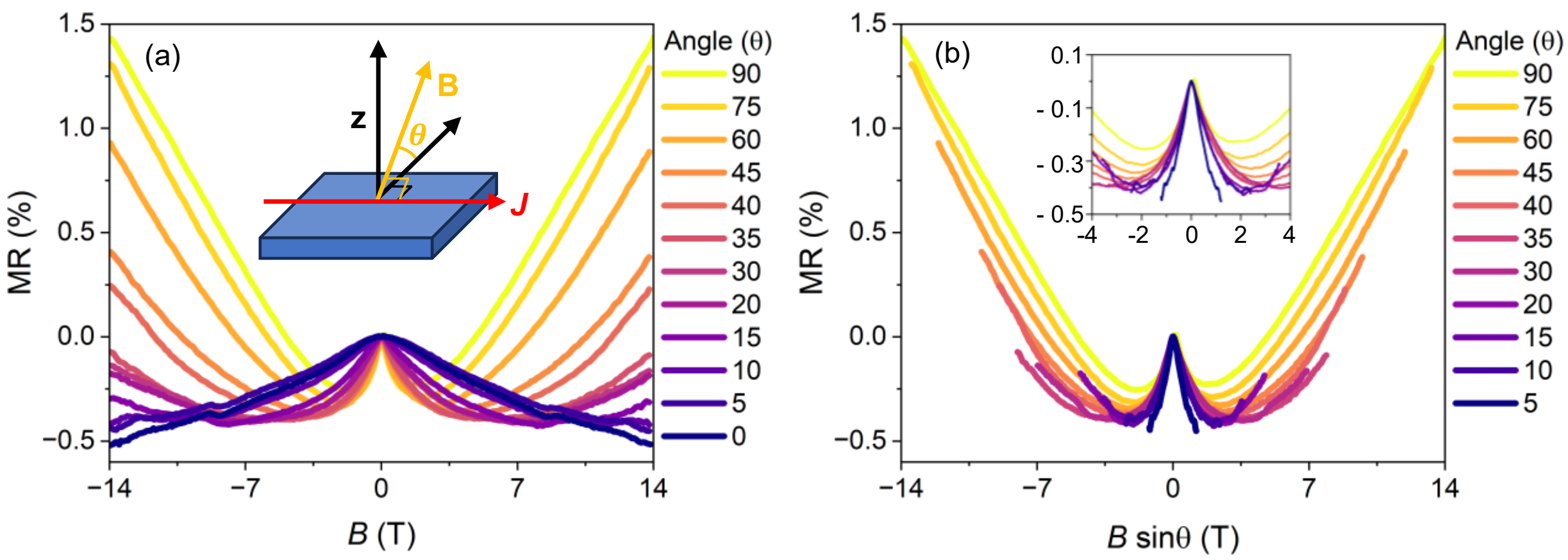

Fig. 3: (a) Angle dependence of the MR of sample $Pr_4Ni_3O_{10}$-S2 at 2 K. Magnetic field and current directions are kept perpendicular to each other (inset). (b) Scaled MR plotted as a function of the perpendicular component $B$ sin(θ). The inset shows the low-field region on expanded scales.

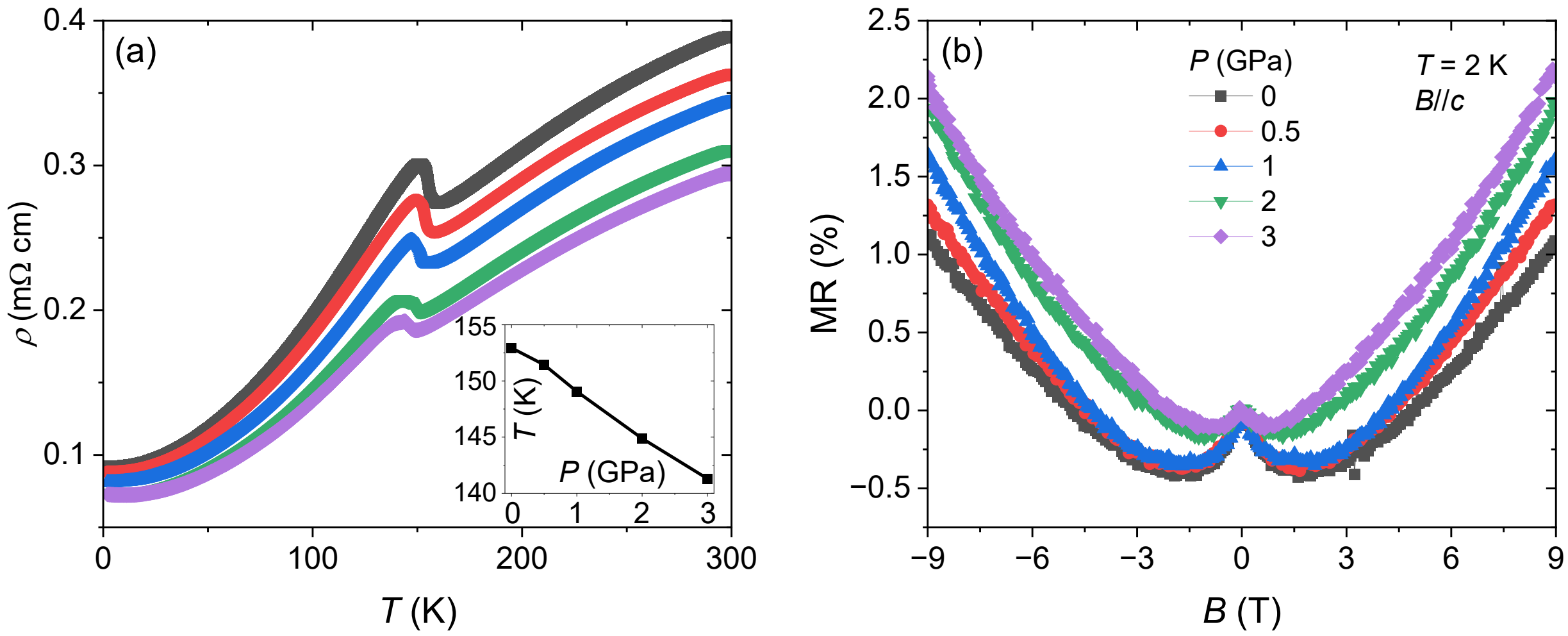

Fig. 4: Temperature dependence of the in-plane resistivity of $Pr_4Ni_3O_{10}$ S3 (a) and magnetoresistance of $Pr_4Ni_3O_{10}$ S3 with $B//c$ under pressure up to 3 GPa (b). The inset in (a) shows the evolution of $T_{DW}$ under pressure.

## Supplemental Materials

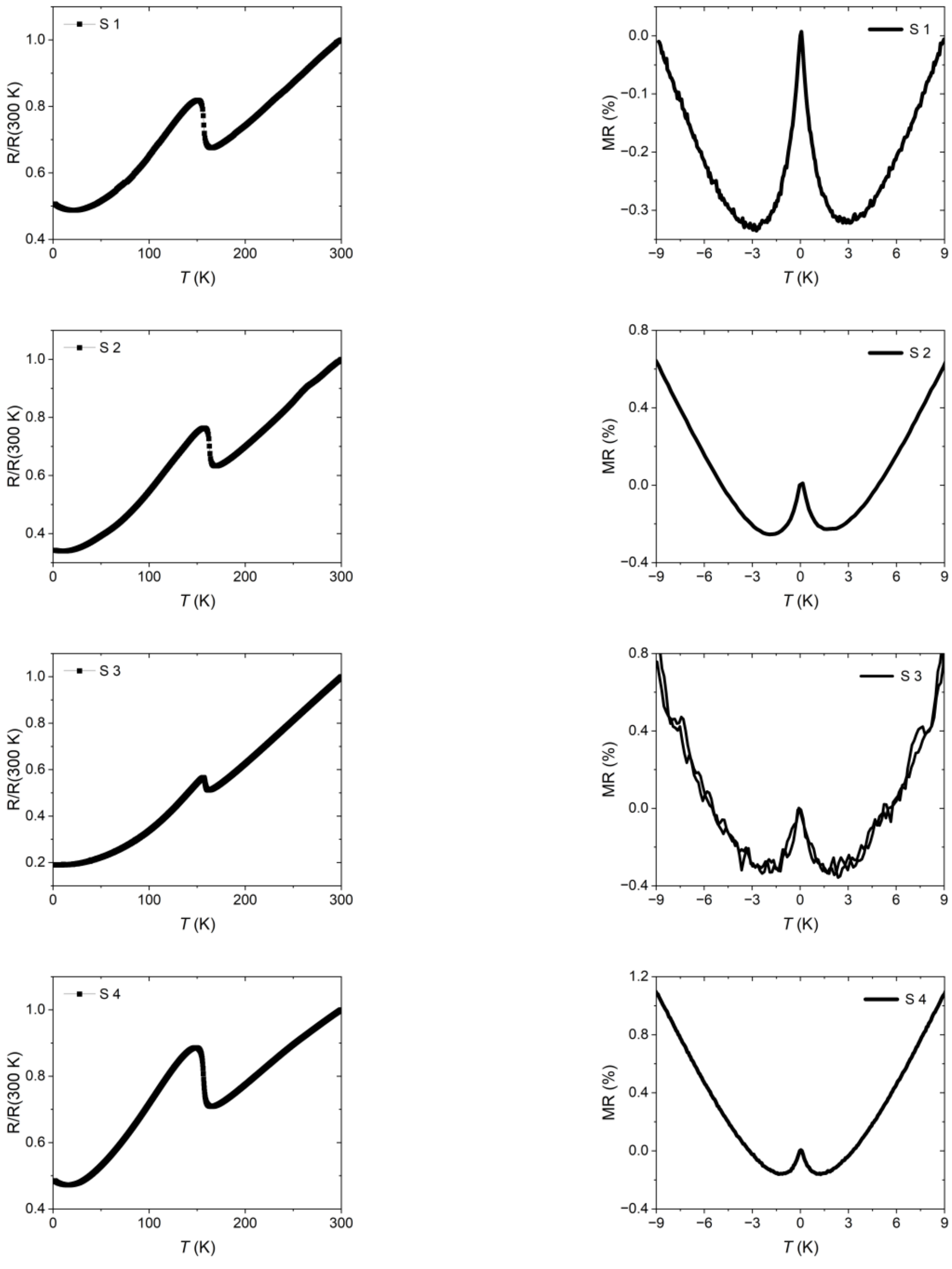

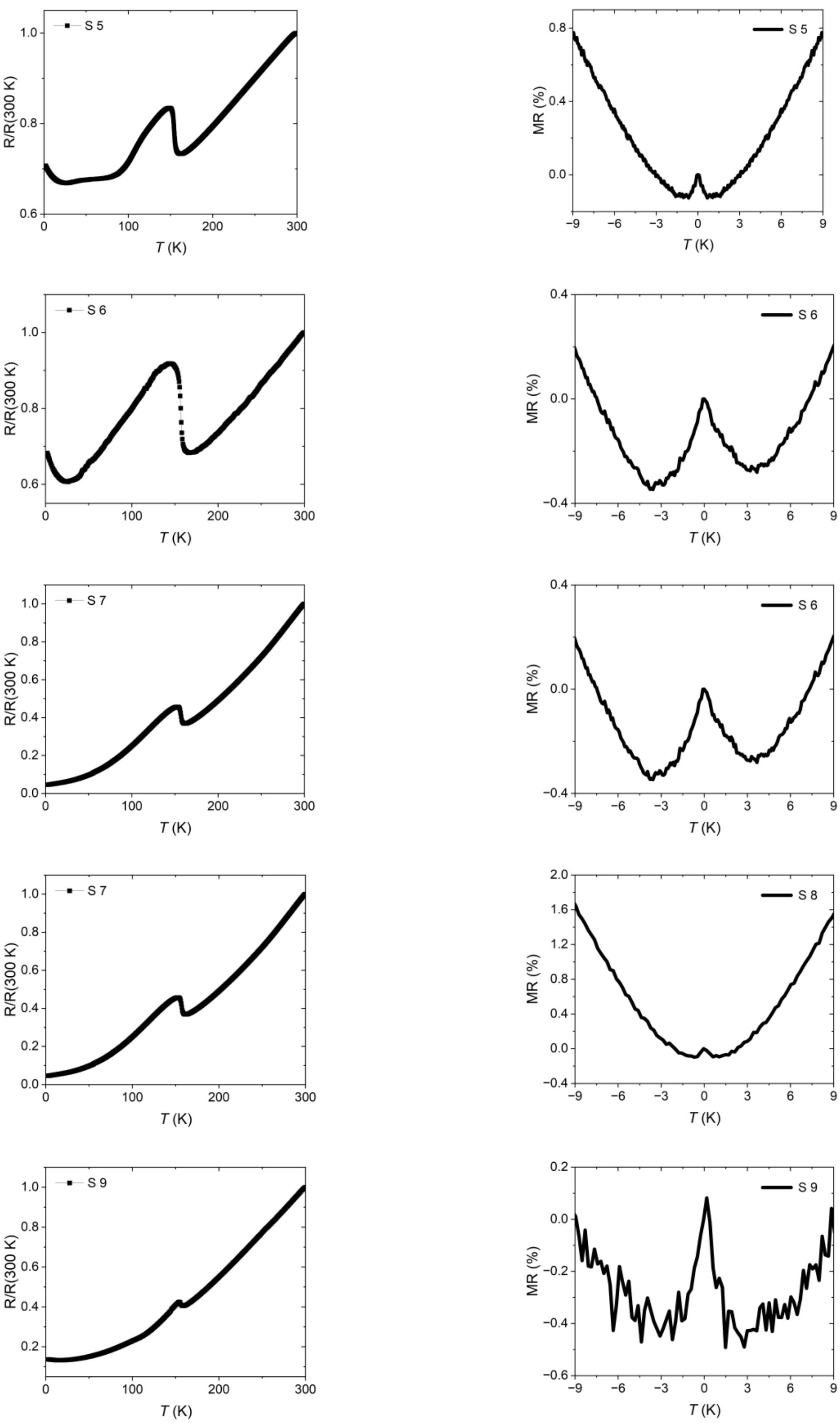
S 5
R/R(300 K)
T (K)
MR (%)
S 6
S 7
S 8
S 9

Fig. S1: Compilation of the temperature dependence of the resistance (plotted as R(T)/R(300K)) (left column) and of the MR at 2 K (right column) of nine PNO crystals. All samples display the signature at the DW transition near 150 K. However, the behavior at low temperatures (saturation or upturn) is sample dependent. Also, all samples show the NMR in low fields. For most of the samples, the NMR reaches ~-0.3 %; a correlation with the low temperature variation of the resistance is not apparent.

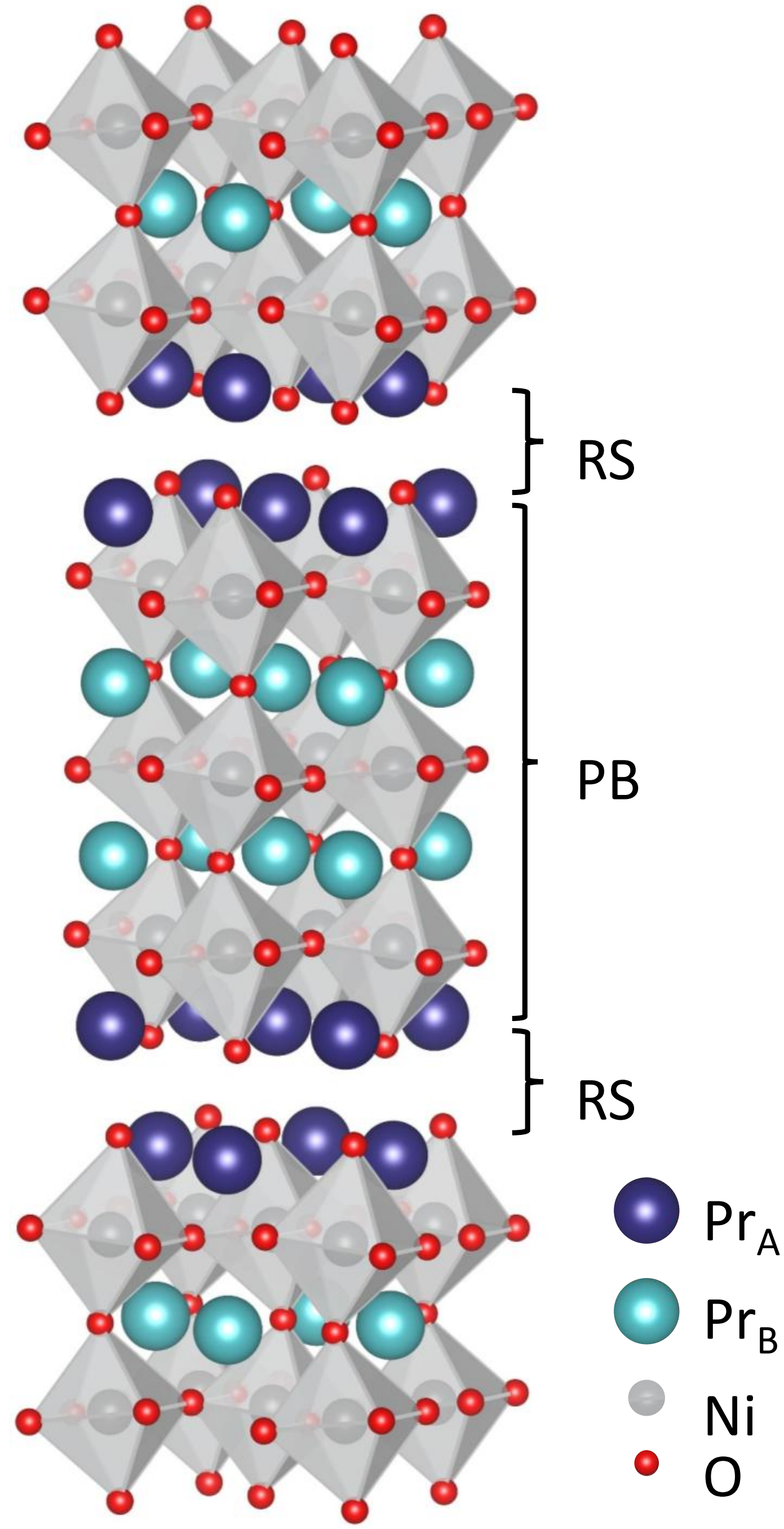

Fig. S2: Crystal structure of $Pr_4Ni_3O_{10}$. RS indicates the rock salt layers and PB the perovskite block. $Pr_A$ and $Pr_B$ denote the two inequivalent rare earth sites.

**Magnetic characterization of the $Pr_4Ni_3O_{10}$ (PNO) and $La_4Ni_3O_{10}$ (LNO) samples**

Figure S3 summarizes the magnetic characteristics of our samples. Fig. S3(a) shows the temperature-dependent susceptibility $\chi=M/H$ of PNO measured in a magnetic field of 2 T applied along the *c*-axis and parallel to the in-plane, respectively. Also included are data for an LNO reference sample. Overall, the susceptibility of LNO amounts is less than ~1% of PNO, a finding in agreement with previous reports [1-3] and reflecting the fact that the $La^{3+}$-ion is non-magnetic. The observed susceptibility of LNO is thus attributed to the Ni ions. In fact, the kink in the temperature dependence of the susceptibility of LNO near 135 K marks the transition into the intertwined charge/spin density wave (CDW/SDW) state residing in the NiO-trilayer block [Zhang-2020]. At low temperatures, the susceptibility of LNO increases rapidly for reasons that have not been yet entirely clarified [3-5]. The temperature dependence of the susceptibility of PNO resembles a Curie-Weiss (CW) behavior, and in the entire temperature range the magnetic anisotropy of $Pr^{3+}$ in PNO between the in-plane and out-of-plane directions is negligible. To reveal the detailed behavior of the magnetism of the $Pr^{3+}$ ions in PNO, we subtract the susceptibility data of LNO from those of PNO and apply a CW fit, $\chi_{CW}(T) = \frac{C}{T-\theta}$,. A fit to the data at temperatures above 200 K yields parameters of $C$ = 4.9 emu mol$^{-1}$ Oe$^{-1}$ K$^{-1}$ and $\theta$ ~ -1 K. The effective magnetic moment derived for PNO is 6.2 $\mu_B$/f.u., similar to previous reports [2, 3, 6], amounting to 3.1 $\mu_B$/Pr, slightly less than the expected full-free-ion moment of $Pr^{3+}$ of 3.58 $\mu_B$. We note though that the susceptibility does not strictly follow the CW form even at higher temperatures as seen in the deviation plot in Fig. S2(b). Therefore, the fit parameters depend on the fitted temperature range and serve merely as estimates [7-9].

At low temperatures, clear deviations from the CW dependence arise as seen by plotting $\Delta 1/\chi = 1/(\chi_{Pr} - \chi_{La}) - 1/\chi_{CW}$ in Fig. S3(b). A step of ~ 0.31x10$^{-4}$ emu/mol Oe near 158 K signals the transition into the CDW/SDW state of PNO. The measured susceptibility falls increasingly short of the CW-fit as the temperature decreases, consistent with induced moment magnetism of non-Kramers ions and the van Vleck scenario [10]. At temperatures below ~25 K, $\Delta\chi$ levels off. This temperature range is close to the onset of magnetic order on the $Pr_A$ sites (see Fig. S2) and possibly related to it, as revealed in a recent resonant x-ray and neutron scattering study [11].

Fig. S3(c) shows isothermal magnetization curves at various temperatures in *c*-axis applied fields. Even in a field of 7 T and at a temperature as low as 2 K, the measured magnetization is far from saturation. This is not expected for a magnetization process following the conventional Brillouin function. Indeed, the data displays a remarkable scaling property when plotted using the variable $H/(T+35\text{K})$ as shown in the inset. This scaling holds over the entire field and temperature ranges except the 2 K data which fall slightly below the trend. In terms of the mean-field formalism, this scaling indicates antiferromagnetic correlations corresponding to an energy scale of ~35K. We note that this scaling is robust and does not depend on details of the CW fit, for instance. Furthermore, there is no obvious deviation in the scaling due to the onset of magnetic order on the Ni-sites at 158 K and the associated temperature-dependent ordered moment. These findings indicate that the antiferromagnetic exchange correlations between Pr ions exist already at high temperatures and can thus promote an induced moment and a magnetically ordered state at low temperatures [12-15] as arises on the $Pr_A$ sites below ~25 K [11].

The temperature-derivative of the susceptibility reveals additional features at low temperatures as shown in the inset of Fig. S3(a). Below ~25 K, $d\chi/dT$ displays an enhanced downwards trend which may be correlated to the onset of magnetic order on the $Pr_A$ sites. Below 10 K, a clearly resolved peak in $d\chi/dT$ appears when measured in a field of 0.5 T. A peak near 5 K was also observed in specific heat measurements [3, 11]; possible causes of crystal field excitations or the onset of magnetic order on the Pr sites in the perovskite block layer ($Pr_B$) have been discussed. The data in Fig. S3(a) also reveal that the low-temperature peak is completely suppressed in a field of 2 T, suggesting that it is magnetic in origin.

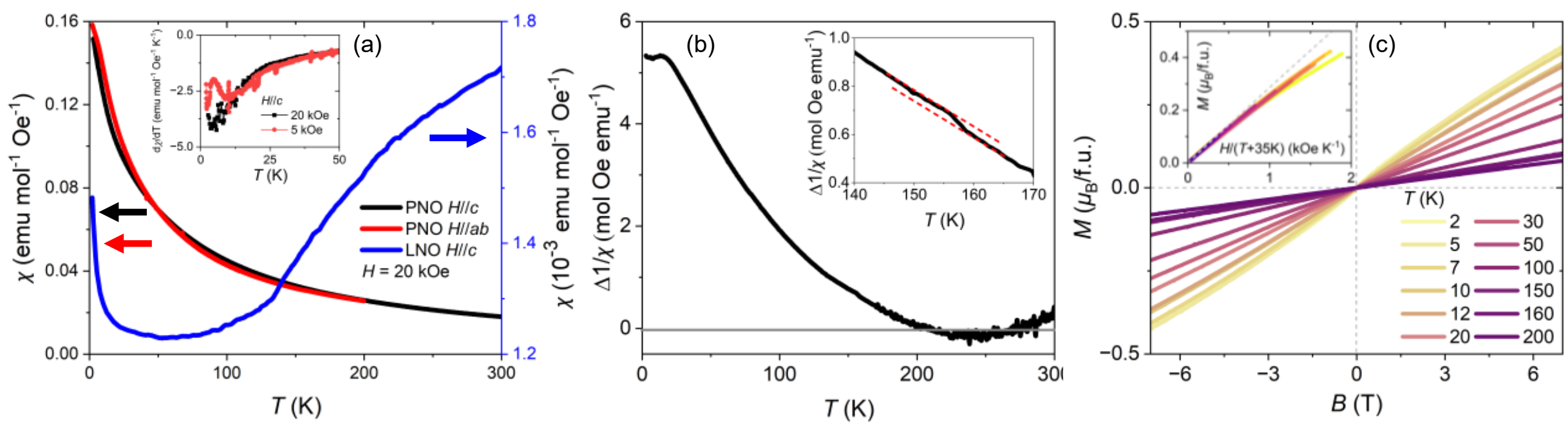

Fig. S3: (a) Temperature dependence of the susceptibility of $Pr_4Ni_3O_{10}$ (left y-axis) and $La_4Ni_3O_{10}$ (right y-axis) measured in an applied field of 2 T. Inset: Temperature derivative of the susceptibility for PNO measured in fields of 0.5 and 2 T. (b) Difference $\Delta\ 1/\chi$ between the measured susceptibility $1/(\chi_{\mathrm{Pr}} - \chi_{\mathrm{La}})$ and the Curie-Weiss fit. The inset highlights the transition into the CDW/SDW state at 158 K. (c) main panel: isothermal magnetization curves of PNO measured in a c-axis field at various temperatures as indicated. Inset: scaling plot of the magnetization in the variable $H/(T$+35K). The dashed line indicates the variation expected from the $J$=4 Brillouin function when plotted as a function of $H/(T$+35K).

**Discussion related to fitting with Eq. 1**

We have attempted to extract the spin–orbit and spin-flip scattering rates independently by fitting the full WL expression, Eq. (1), without the simplifying approximations. This relation contains three parameters, $H_\varphi$*, $H_{so}$* and the overall amplitude, $\alpha$. As explained in the manuscript, a fourth parameter, $\beta$, accounts for the Drude-type background magnetoresistance. It turns out that it is not possible to obtain reliable fit values for $H_{so}$*, presumably because the contributions from the spin–orbit and spin-flip scattering terms are much smaller than that of the inelastic scattering term. For example, at 2 K, the extracted fitting parameters are $H_\varphi^* = 0.0924 \pm 0.01705$ and $H_{so}^* = -0.04167 \pm 0.02045$ with a dependency of 0.9996. We note that the obtained parameter values satisfy the condition of $H_{so}^*/H_\varphi^* < 9/16$. Furthermore, the negative value of $H_{so}^*$ is the expected result obtained from the in-plane data in Fig. 3(a). However, the large error bars, in particular of $H_{so}^*$ of ~50 %, in conjunction with their mutual dependency of almost 1 indicated to us that the obtained fitting parameters are quite unreliable. We remark that a dependency of close to 1 indicates that distinct fit parameters cannot be determined as different combinations yield fits of similar quality. We therefore chose not to include separate fits in the manuscript. Also, since $H_\varphi^* = H_\varphi + 2H_s$ and $H_{so}^* = H_{so} - H_s$, the bare values of $H_s$ and $H_{so}$ cannot be determined.